\documentclass[aps,pre,twocolumn,groupedaddress,10pt,showkeys]{revtex4}

\newcommand{\rv}{{\bf r}}
\newcommand{\Trcl}{\mathrm{Tr_{cl}}}
\newcommand{\Tr}{\mathrm{Tr}}
\newcommand{\Fcal}{\mathcal{F}}
\begin{document}

\title{Variational Principle of Classical Density Functional Theory
  via Levy's Constrained Search Method}

\author{Wipsar Sunu Brams Dwandaru}
\affiliation{H. H. Physics Laboratory, University of Bristol, Royal
 Fort, Tyndall Avenue, Bristol, BS8 1TL, UK} 
\affiliation{Jurusan Fisika, Universitas Negeri Yogyakarta,
  Bulaksumur, Yogyakarta, Indonesia}

\author{Matthias Schmidt}
\affiliation{H. H. Physics Laboratory, University of Bristol, Royal
  Fort, Tyndall Avenue, Bristol, BS8 1TL, UK}
\affiliation{Theoretische Physik II, Physikalisches Institut,
  Universit\"{a}t Bayreuth, Universit\"{a}tsstra\ss e, D-95440
  Bayreuth, Germany}

\date{15 April 2011, to appear in Phys. Rev. E}

\begin{abstract}
  We show that classical density functional theory can be based on the
  constrained search method [M. Levy, {\em Proc. Natl. Acad. Sci.}
    {\bf 76}, 6062 (1979)].  From the Gibbs inequality one first
  derives a variational principle for the grand potential as a
  functional of a trial many-body distribution. This functional is
  minimized in two stages. The first step consists of a constrained
  search of all many-body distributions that generate a given one-body
  density. The result can be split into internal and external
  contributions to the total grand potential. In contrast to the
  original approach by Mermin and Evans, here the intrinsic Helmholtz
  free energy functional is defined by an explicit expression that
  does not refer to an external potential in order to generate the
  given one-body density. The second step consists of minimizing with
  respect to the one-body density. We show that this framework can be
  applied in a straightforward way to the canonical ensemble.
\end{abstract}


\maketitle

\section{Introduction}

The variational principle of density functional theory (DFT) was
originally formulated for ground-state properties of quantum systems
by Hohenberg and Kohn in 1964 \cite{Hohenberg}. The extension to
non-zero temperatures was performed by Mermin in the following
year~\cite{Mermin}, here still formulated for quantum systems. The
application to the statistical mechanics of classical systems,
i.e.\ the development of classical DFT, was initiated about a decade
later through the work of Ebner, Saam, and Stroud~\cite{Ebner, Ebner2,
  Ebner3}. The generality of the framework was fully realized by
Evans~\cite{Evans3}. His 1979 article continues to be the standard
reference on the subject; there are more recent review \cite{Evans2,
  Evans3, Evans4, Roth} and textbook \cite{Hansen} presentations.

The Hohenberg-Kohn theorem applies to one-particle density
distributions that correspond to a particular external (one-body)
potential $v$ in the Hamiltonian, in which the kinetic and internal
interaction are those of the true system \cite{Hohenberg, Mermin}. One
refers to $v$-representability of the one-particle density, i.e.\ the
condition that a one-particle density is generated by some external
potential $v$. However, it was realized, already in the original
Hohenberg-Kohn paper, that $v$-representability is not guaranteed for
an arbitrarily chosen density~$\rho$ \cite{Hohenberg, Levy,
  Levy2}. One argues that this does not pose any problems in the
practical applications of DFT to quantum systems~\cite{Kohn}. In the
development of the theory, it turned out that there indeed exist
non-$v$-representable densities, i.e.\ one-body densities that are not
associated to any ground-state wave function~\cite{Levy3}.  The
original Hohenberg-Kohn theorem does not apply to these.

In 1979 Levy introduced an alternative foundation of DFT for quantum
systems, based on a constrained, two-stage search~\cite{Levy}. Here a
weaker condition, known as $N$-representability, is used, where the
density distribution may be directly obtained from some anti-symmetric
$N$-body wave function, although an external potential that generates
this wave function need not exist~\cite{Levy, Levy3, Coleman}. One
defines an exchange-correlation functional that demands searching all
wave functions that return the fixed (trial) one-body density. The
latter need not be $v$-representable. Subsequently, a method similar
to Levy's was proposed by Lieb \cite{Lieb}, called the generalized
Legendre transform~\cite{Levy2}. Instead of searching all wave
functions, the functional searches all possible external potentials
that correspond to a fixed density. Kohn adopted the constrained
search for his Nobel lecture~\cite{Kohn}, and it is viewed as an
important theoretical contribution to the foundation of DFT for
electronic structure. Practical applications of constrained search
functionals are of ongoing research interest, see
e.g.~\cite{Levy2,Levy4}. Levy gives a brief historic account of the
development of his ideas in Ref.\ \cite{levy2010}.

Given the significance of Levy's and Lieb's methods for electronic
structure, it is somewhat surprising that there are very few studies
that point to the use of these in classical systems. One example is
the work by Weeks~\cite{Weeks}, where the $v$-representability of the
one-body density in some finite region of space is investigated
through the Gibbs inequality. Although Weeks cites Levy's original
paper~\cite{Levy}, and makes a remark that his formulation is in
spirit similar to that of Levy, it seems that his method is related
more to Lieb's generalized Legendre transform method. Earlier work has
been carried out in order to investigate the existence of an external
potential that is associated to a given equilibrium one-body
density~\cite{Chayes}. It is concluded that there is such an external
potential that produces any given density for any (classical) system
without hard core interaction. Although one might guess from general
arguments that the constrained search can be applied to classical DFT,
to the best of our knowledge, this procedure has not been spelled out
explicitly in the literature.

In the present article we show how to formulate the variational
principle of classical DFT based on Levy's constrained search method.
This alternative can provide further insights into the foundation of
classical DFT. In particular the intrinsic free energy functional is
defined here without implicit reference to an external potential $v$.
A more relaxed condition for $\rho$, similar to that of
$N$-representability, is imposed. Here, the one-body density is only
required to be obtained from an arbitrarily chosen many-body
probability distribution $f$. We refer to this condition as
$f$-representability of a given $\rho$.  While distinguishing between
the different type of representability in practical DFT calculations
seems unnecessary, we find the discrimination very useful for
conceptual purposes and hence point out throughout the manuscript
which of the conventions is followed in the reasoning.  We also show
that Levy's method can be applied in a straightforward way to the
canonical ensemble.  There is considerable current interest in the
theoretical description of the behaviour of small systems, where the
grand and the canonical ensembles are inequivalent in general, and the
later might model certain (experimental) realizations of strongly
confined systems more closely. Several relatively recent contributions
address the problem of formulating DFT in the canonical ensemble
\cite{White00prl,White00pre,Hernando02,Gonzalez04jcp,White02,HernandoBlum02}.
The authors of these papers consider the important problem of how to
obtain DFT approximations that make computations in the canonical
ensemble feasible. Our present article has a much lower goal: We are
only concerned with formulating the variational principle in an
alternative way.

This article is organized as follows. We start by defining the grand
potential as a functional of the many-body probability distribution in
Sec.~\ref{SEComegaManyBody}. This is a necessary step and our
presentation follows ~\cite{Evans2, Evans3, Evans4, Hansen}. In
Sec.~\ref{SECmerminEvans} we give a brief overview of the standard
proof of DFT, expressing the free energy as a functional of the
one-body density based on a one-to-one correspondence between the
one-body density and the external potential. The full derivation is
widely known and can be found in numerous references~\cite{Evans2,
  Evans3, Evans4, Hansen, Widom2}. We proceed, in Sec.\ \ref{SEClevy},
by formulating the intrinsic free energy functional via the
constrained search method; our presentation is similar to Levy's
original work~\cite{Levy}. Our central result is the definition
(\ref{EQFviaLevy}) of the intrinsic free energy functional, without
reference to an external potential. We summarize the essence of Levy's
argument as a double minimization \cite{levy2010} in
Sec.\ \ref{SECtwoStageMinimization}. In Sec.\ \ref{SECcanonical} we
apply this to the canonical ensemble and we conclude in
Sec.\ \ref{SECconclusions}.

\section{Grand potential functional of the many-body distribution}
\label{SEComegaManyBody}
In the grand canonical ensemble of a system of classical particles,
the equilibrium probability distribution for $N$ particles at
temperature $T$ is assumed to exist and to be given by
\begin{equation}
  f_0=\Xi^{-1}\exp\left(-\beta\left(H_{N} - \mu N\right)\right),
  \label{EQf0}
\end{equation}
where $H_{N}$ is the Hamiltonian of $N$ particles, $\mu$ is the
chemical potential, and $\beta = 1/(k_{B}T)$, with $k_{B}$ being the
Boltzmann constant. The normalization constant is the grand canonical
partition sum
\begin{equation}
  \Xi = \mathrm{Tr_{cl}}\exp\left(-\beta\left(H_{N}-\mu N\right)\right),
\end{equation}
where $\mathrm{Tr_{cl}}$ represents the classical trace, i.e.\ the sum
over total particle number and integral over all degrees of freedom
\begin{equation}
  \mathrm{Tr_{cl}} = \sum^\infty_{N=0}
  \frac{1}{h^{3N}N!}\int\mathrm{d}\rv_1\ldots\mathrm{d}\rv_N
  \int\mathrm{d}{\bf p}_1\ldots\mathrm{d}{\bf p}_N,
\end{equation}
where $h$ is the Planck constant, $\rv_{1}, \ldots, \rv_{N}$ are the
position coordinates and ${\bf p}_{1}, \ldots, {\bf p}_{N}$ are the
momenta of particles $1, \ldots, N$.

One introduces the grand potential as a functional of the many-body
probability distribution,
\begin{equation}
  \Omega[f] = \mathrm{Tr_{cl}}f\left(H_{N}-\mu N + \beta^{-1}\ln f\right),
  \label{EQomegaMermin}
\end{equation}
where $f$ is a variable trial probability distribution that satisfies
the normalization condition
\begin{equation}
  \mathrm{Tr_{cl}}f = 1.
  \label{EQfNormalized}
\end{equation}
Note that $f$ as an argument of functional (\ref{EQomegaMermin}) can
be quite general and need not be linked to an external potential at
this stage. Inserting the equilibrium probability
distribution~(\ref{EQf0}) into (\ref{EQomegaMermin}) one obtains
\begin{eqnarray}
  \Omega[f_0]&=&
  \Trcl f_0\left(H_N-\mu N+\beta^{-1}\ln f_0\right)\nonumber\\
  &=&\Trcl f_0\left(H_N-\mu N+\beta^{-1}
      \left[-\ln\Xi-\beta(H_N-\mu N)\right]\right)\nonumber\\
  &=&-\beta^{-1}\ln\Xi\equiv\Omega_0,
      \label{EQomega0}
\end{eqnarray}
where $\Omega_0$ is the equilibrium grand potential. An important
property of functional (\ref{EQomegaMermin}) is that it satisfies the
variational principle
\begin{equation}
  \Omega[f] > \Omega[f_0], \quad f\neq f_0,
  \label{EQmanyBodyVariational}
\end{equation}
which may be proven using the Gibbs-Bogoliubov relation as
follows. First, from (\ref{EQf0}) and (\ref{EQomega0}), $\Omega[f]$ of
Eq.\ (\ref{EQomegaMermin}) can be written as
\begin{equation}
  \Omega[f] = \Omega[f_0] + 
    \beta^{-1}\Trcl f\ln\left(\frac{f}{f_0}\right).
    \label{EQomegaPreInequality}
\end{equation} 
According to the Gibbs inequality~\cite{Evans3,Hansen},
\begin{equation}
  f\ln\left(\frac{f_0}{f}\right) < f\left(\frac{f_0}{f} - 1\right),
\end{equation}
and hence
\begin{equation}
  \Trcl f\ln\left(\frac{f}{f_0}\right)>\Trcl(f-f_0).\nonumber
\end{equation}
Since $f$ and $f_0$ are normalized, i.e.\ satisfy
(\ref{EQfNormalized}), the RHS of the inequality above vanishes, and
\begin{equation}
  \beta^{-1}\Trcl f\ln\left(\frac{f}{f_0}\right) > 0.
\end{equation}
Thus the second term on the RHS of (\ref{EQomegaPreInequality}) is
positive, and the inequality (\ref{EQmanyBodyVariational}) follows.

For classical particles the Hamiltonian may be restricted to the form
\begin{equation}\label{EQhamiltonian}
  H_N = \sum^N_{i=1}\frac{p_i^2}{2m} +
  U(\rv_1,\ldots,\rv_N)+\sum^N_{i=1}v(\rv_i),
\end{equation}
where the first term is the total kinetic energy, with the squared
momentum $p_i^2={\bf p}_i\cdot {\bf p}_i$ of the $i$-th particle,
$U$~is the interatomic potential between the particles, $v$ is an
(arbitrary) external one-body potential and $m$ is the particle
mass. The equilibrium one-body density at position $\rv$ is given as a
configurational average
\begin{equation}
  \rho_0(\rv) = \Trcl f_0 \hat\rho(\rv),
  \label{EQrho0definition}
\end{equation}
where the density operator for $N$ particles is defined as
\begin{equation}
  \hat\rho(\rv) = \sum^N_{i=1}\delta(\rv-\rv_i).
  \label{EQdensityOperator}
\end{equation}

The functional form (\ref{EQomegaMermin}) was originally introduced by
Mermin~\cite{Mermin} for (finite temperature) quantum systems, where
the grand potential is a functional of a (trial) density matrix. The
variational principle (\ref{EQmanyBodyVariational}) will be used in
Secs.\ \ref{SECmerminEvans} and \ref{SEClevy} below, where we present
two alternative derivations of the intrinsic free energy as a
functional of the one-body density.

\section{Mermin-Evans Derivation of the Free Energy Functional}
\label{SECmerminEvans}
Evans gave a formal proof that the intrinsic free energy of a system
of classical particles is a functional of the one-body density
\cite{Evans2}. Here we briefly lay out his chain of arguments. The
many-body distribution $f_0$ as given in (\ref{EQf0}) is a functional
of the external potential $v$ through Eq.\ (\ref{EQhamiltonian}), and
therefore $\rho_0$ is a functional of $v$ via
(\ref{EQrho0definition}). This, in principle, requires solution of the
many-body problem and the dependence is in accordance with physical
intuition, i.e.\ it is the action of $v$ that generates the shape of
the density profiles $\rho_0$.

However, the more useful result that can be deduced~\cite{Evans2}, is
that once the interatomic interaction potential $U$ is given, $f_0$ is
a functional of $\rho_0$. The proof of this statement rests on
reductio ad absurdum~\cite{Evans2,Hansen}, where for a given
interaction potential $U$, $v$ is uniquely determined by $\rho_0$. The
resultant $v$ then determines $f_0$ via (\ref{EQf0}) and
(\ref{EQhamiltonian}). Hence, $f_0$ is a functional of $\rho_0$.

An important consequence in this reasoning is that for given
interaction potential $U$,
\begin{equation}
  \Fcal[\rho] = \Trcl f_0\left(
    \sum^N_{i=1}\frac{p^2_i}{2m} + U + \beta^{-1}\ln f_0\right)
    \label{EQFviaMerminEvans}
\end{equation}
is a unique functional of the (trial) one-body density $\rho$. Here,
the dependence of $f_0$ on the external potential, $v$, is now only
implicit through the one-body density, $\rho$. We will comment on this
sequence of dependences in the conclusions, after having laid out
Levy's alternative method to define a free energy functional in
Sec.\ \ref{SEClevy}. Furthermore, using a Legendre transform, the
grand potential functional is obtained for a given external potential
as
\begin{equation}
  \Omega_v[\rho] = \Fcal[\rho] + 
  \int\mathrm{d}\rv(v(\rv)-\mu)\rho(\rv).
\end{equation}
The functional $\Omega_v[\rho]$ returns its minimum value if $\rho =
\rho_0$, i.e.\ if the trial density is the true equilibrium one-body
density of the system under the influence of $v$. The value is the
grand potential $\Omega_0$. The existence of the minimum value of
$\Omega_v[\rho]$ may be proven by considering another equilibrium
density $\rho'$ associated with a probability distribution $f'$ of
unit trace, such that
\begin{eqnarray}
  \Omega[f']&=&\Trcl f'\left(H_N-\mu N + \beta^{-1}\ln f'\right)\nonumber\\
  &=&\Fcal[\rho']
  +\int\mathrm{d}\rv\left(v(\rv)-\mu\right)\rho'(\rv)\\
  &=&\Omega_v[\rho'],\nonumber
\end{eqnarray}
where
\begin{equation}
  \Fcal[\rho'] = \Trcl f'\left(
  \sum^N_{i=1}\frac{p^2_i}{2m} + U + \beta^{-1}\ln f'\right).
\end{equation}
However, it is known from Eq.\ (\ref{EQmanyBodyVariational}) that
$\Omega[f']>\Omega[f_0]$, for $f'\neq f_0$, thus it follows that
\begin{equation}
  \Omega_{v}[\rho']>\Omega_{v}[\rho_0].
\end{equation}
In other words, the correct equilibrium one-body density, $\rho_0$,
minimizes $\Omega_v[\rho]$ over all density functions that can be
associated with a potential $v$.

This important result may be stated as a functional derivative
\begin{equation}
\left.
\frac{\delta\Omega_v[\rho]}{\delta\rho(\rv)}\right|_{\rho_0}=0,
\end{equation}
and
\begin{equation}
  \Omega_{v}[\rho_0] = \Omega_0.
\end{equation}

To conclude, the formal argument for the definition
(\ref{EQFviaMerminEvans}) of the intrinsic free energy functional,
$\Fcal[\rho]$, is based on $v$-representability of the one-body
density. A $v$-representable $\rho$ is one which is associated with a
probability distribution, $f$, of the given Hamiltonian $H_{N}$ with
external potential $v$~\cite{Levy, Kohn}.  This condition was
originally introduced for quantum systems, and is implicit in the
current approach. It is used to prove the chain of dependency outlined
above (\ref{EQFviaMerminEvans}), and confirmed for a large class of
(classical) systems in~\cite{Chayes}.

\section{Free energy functional via Levy's constrained search method}
\label{SEClevy}
Here we show how one may alternatively define a free energy functional
via Levy's method. This is based on the weaker condition of
$f$-representability, where trial density fields $\rho$ need not
necessarily be associated with some external potential. We define the
intrinsic Helmholtz free energy functional as
\begin{equation}
  \Fcal_{\mathrm{L}}[\rho] = 
  \min_{f\rightarrow\rho}
  \left[\Trcl f\left(\sum^N_{i=1}\frac{p^{2}_{i}}{2m}+U+ 
    \beta^{-1}\ln f\right)\right],
  \label{EQFviaLevy}
\end{equation}
where the minimization searches all probability distributions $f$,
that are normalized according to (\ref{EQfNormalized}), and that yield
the fixed trial one-body density $\rho$ via
\begin{equation}
  \rho(\rv) = \Trcl f\hat\rho(\rv).
  \label{EQfTorho}
\end{equation}
The notation $f\to\rho$ in (\ref{EQFviaLevy}) indicates the
relationship (\ref{EQfTorho}). Note that i) in general there will be
many different forms of $f$ that yield the same $\rho$, and ii) no
further conditions on~$f$ are imposed, apart from its
normalization. In particular, the form of $f$ need not be of
Boltzmann-type containing the interaction potential $U$ (as was the
case in Sec.\ \ref{SECmerminEvans}). Hence~$\rho$ need only be
$f$-representable, but not necessarily
$v$-representable. $\Fcal_{\mathrm{L}}[\rho]$ returns a minimum value
by choosing the probability distribution that minimizes the term in
brackets in (\ref{EQFviaLevy}). Note that the functional form of this
term is formally equivalent to (\ref{EQFviaMerminEvans}) and that it
is a sum of contributions due to kinetic energy, internal interaction
energy $U$, and (negative) entropy $k_B f\ln f$ multiplied by $T$.

The grand potential functional for a given external potential is then
\begin{equation}
  \Omega_{\mathrm{L}}[\rho] = 
  \Fcal_{\mathrm{L}}[\rho] +
  \int\mathrm{d}\rv\left(v(\rv)-\mu\right)\rho(\rv).
  \label{EQomegaLevy}
 \end{equation}
This functional possesses two important properties. i) At the
equilibrium density it yields the equilibrium grand potential
\begin{equation}
  \Omega_{\mathrm{L}}[\rho_0] = \Omega_0,
  \label{EQomega0Levy}
\end{equation}
where $\rho_0$ is given by (\ref{EQrho0definition}) and $\Omega_0$ by
(\ref{EQomega0}). This value also constitutes the minimum such that
\begin{equation}
  \Omega_{\mathrm{L}}[\rho]\geq\Omega_0.
  \label{EQomegaLevyVariational}
\end{equation}

In order to prove (\ref{EQomega0Levy}) and
(\ref{EQomegaLevyVariational}), we introduce additional notation. Let
$f^\rho_\mathrm{min}$ be the probability distribution that satisfies
the RHS of Eq.\ (\ref{EQFviaLevy}). Then it follows that
\begin{equation}
  \Fcal_{\mathrm{L}}[\rho] =
  \Trcl f^\rho_\mathrm{min}
  \left(\sum^N_{i=1}\frac{p^2_i}{2m} + U + 
  \beta^{-1}\ln f^\rho_\mathrm{min}\right),
\end{equation}
and for the case of the equilibrium density
\begin{equation}
  \Fcal_\mathrm{L}[\rho_0] = 
  \Trcl f^{\rho_0}_\mathrm{min}\left(
  \sum^N_{i=1}\frac{p^2_i}{2m} + U + 
  \beta^{-1}\ln f^{\rho_0}_\mathrm{min}\right).
  \label{EQFLalternative}
\end{equation}

First we proof the inequality (\ref{EQomegaLevyVariational}). By its
very definition (\ref{EQomegaLevy}), the LHS of
(\ref{EQomegaLevyVariational}) may be rearranged into
\begin{eqnarray}\label{2.88}
  &&\hspace{-8mm}\int\mathrm{d}\rv\left(v(\rv)-\mu\right)\rho(\rv)
  +\Fcal_{\mathrm{L}}[\rho]\nonumber\\
  &&=\int\mathrm{d}\rv\left(v(\rv)-\mu\right)\rho(\rv)\nonumber\\
  &&  \quad + \Trcl f^\rho_\mathrm{min}
  \left(\sum^{N}_{i=1}\frac{p^{2}_{i}}{2m} 
  + U + \beta^{-1}\ln f^\rho_{\mathrm{min}}\right)\\
  &&=\Trcl f^\rho_\mathrm{min}\left(H_N -\mu N 
  +\beta^{-1}\ln f^\rho_\mathrm{min}\right).\nonumber
\end{eqnarray}
But according to the inequality (\ref{EQmanyBodyVariational}),
\begin{equation}\label{2.89}
  \Trcl f^{\rho}_{\mathrm{min}}
  \left(H_N - \mu N + \beta^{-1}\ln f^\rho_\mathrm{min}\right)\geq\Omega_0.
\end{equation}
Thus, combining (\ref{2.88}) and (\ref{2.89}), the inequality
(\ref{EQomegaLevyVariational}) is recovered. In order to prove
(\ref{EQomega0Levy}), it is obvious from (\ref{EQmanyBodyVariational})
that
\begin{equation}
  \Trcl f^{\rho_0}_\mathrm{min}\left(H_{N} - \mu N 
  + \beta^{-1}\ln f^{\rho_0}_\mathrm{min}\right)\geq\Omega_0,
\end{equation}
or, recalling (\ref{EQomega0}),
\begin{eqnarray}
  &&\Trcl f^{\rho_0}_\mathrm{min}
  \left(H_N - \mu N + \beta^{-1}\ln f^{\rho_0}_\mathrm{min}\right)\nonumber\\
  &&\quad\geq\Trcl f_0\left(H_N- \mu N + \beta^{-1}\ln f_0\right).
\end{eqnarray}
But $f^{\rho_0}_{\mathrm{min}}$ and $f_0$ generate the same one-body
density $\rho_0$, hence from
\begin{eqnarray}
  &\mbox{}&\int\mathrm{d}\rv\left(v(\rv)-\mu\right)\rho_0(\rv)\nonumber\\
  &\mbox{}&\quad+\Trcl f^{\rho_0}_{\mathrm{min}}
  \left(\sum^N_{i=1}\frac{p^2_i}{2m} + U 
  + \beta^{-1}\ln f^{\rho_0}_\mathrm{min}\right)\geq\nonumber\\
  &\mbox{}&\int\mathrm{d}\rv\left(v(\rv) - \mu\right)\rho_0(\rv)\nonumber\\
  &\mbox{}&\quad + \Trcl f_0\left(\sum^N_{i=1}\frac{p^2_i}{2m} 
  + U + \beta^{-1}\ln f_0\right),
\end{eqnarray}
we obtain
\begin{eqnarray}
  &&\Trcl f^{\rho_0}_{\mathrm{min}}
  \left(\sum^N_{i=1}\frac{p^2_i}{2m} + U 
  + \beta^{-1}\ln f^{\rho_0}_\mathrm{min}\right)\geq\nonumber\\
  &&\Trcl f_0\left(\sum^{N}_{i=1}\frac{p^2_i}{2m} 
  + U + \beta^{-1}\ln f_0\right).
\end{eqnarray}
However, by the very definition of $f^{\rho_0}_\mathrm{min}$, the
following inequality should also hold:
\begin{eqnarray}
  &&\Trcl f^{\rho_0}_\mathrm{min}
  \left(\sum^N_{i=1}\frac{p^2_i}{2m} + U 
  + \beta^{-1}\ln f^{\rho_0}_\mathrm{min}\right)\leq\nonumber\\
  &&\Trcl f_0\left(\sum^N_{i=1}\frac{p^2_i}{2m} 
  + U + \beta^{-1}\ln f_0\right).
\end{eqnarray}
The above two inequalities hold simultaneously, if and only if
equality is attained,
\begin{eqnarray}\label{2.95}
  &\mbox{}&\Trcl f^{\rho_0}_\mathrm{min}
  \left(\sum^N_{i=1}\frac{p^2_i}{2m} + U 
  + \beta^{-1}\ln f^{\rho_0}_\mathrm{min}\right)\nonumber\\
  &\mbox{}& =  \Trcl f_0\left(\sum^N_{i=1}
  \frac{p^2_i}{2m} + U + \beta^{-1}\ln f_0\right).
\end{eqnarray}
Inserting (\ref{EQFLalternative}) into (\ref{2.95}) yields
\begin{equation}\label{2.96}
  \Fcal_\mathrm{L}[\rho_0] = \Trcl f_0\left(\sum^{N}_{i=1}
  \frac{p^2_i}{2m} + U + \beta^{-1}\ln f_0\right).
\end{equation}
Furthermore, as
\begin{eqnarray}\label{2.97}
\Omega_0&=&\Trcl f_0\left(H_N - \mu N 
+ \beta^{-1}\ln f_0\right)\nonumber\\
&=&\int\mathrm{d}\rv\left(v(\rv) - \mu\right)\rho_0(\rv)\nonumber\\
&& + \Trcl f_0\left(\sum^N_{i=1}\frac{p^2_i}{2m} + U + \beta^{-1}\ln f_0\right),
\end{eqnarray}
inserting (\ref{2.96}) into (\ref{2.97}) returns (\ref{EQomega0Levy}),
which completes the proof. Equation (\ref{2.96}) implies that if
$\rho$ is $v$-representable, then $\Fcal_{\mathrm{L}}[\rho] =
\Fcal[\rho]$. Moreover, $f_0 = f^{\rho_0}_{\mathrm{min}}$ means that
$f_0$ may be obtained directly from $\rho_0$ even if $v$ is unknown:
find the probability distribution which yields $\rho_0$ and which
minimizes (\ref{EQFviaLevy}).

Finally, the inequality (\ref{EQomegaLevyVariational}) implies that
the functional derivative of the grand potential functional vanishes
at equilibrium,
\begin{equation}\label{2.97b}
\left.\frac{\delta\Omega_{\mathrm{L}}[\rho]}
     {\delta\rho(\rv)}\right|_{\rho_0} = 0.
\end{equation}

For completeness we mention that it is convenient to split
$\Fcal_{\mathrm{L}}[\rho]$ into two terms, viz.\ the ideal and excess
free energy functionals, $\Fcal_\mathrm{id}[\rho]$ and
$\Fcal_\mathrm{exc}[\rho]$, respectively, such that
\begin{equation}\label{2.78}
  \Fcal_\mathrm{exc}[\rho] \equiv
  \Fcal_\mathrm{L}[\rho] - \Fcal_\mathrm{id}[\rho],
\end{equation}
where the free energy of the ideal gas (with no interaction potential
present, $U = 0$) is given by
\begin{equation}\label{2.79}
  \Fcal_\mathrm{id}[\rho] = \beta^{-1}\int\mathrm{d}\rv\rho(\rv)
  \left(\ln\left(\lambda^3\rho(\rv)\right) - 1\right),
\end{equation}
where $\lambda = \left(h^{2}\beta/(2m\pi)\right)^{1/2}$.
Thermodynamics enters by realizing that $\Fcal_{\mathrm{L}}[\rho_0]$
is the `intrinsic' Helmholtz free energy of the system, such that the
total free energy is the sum of internal and external contributions,
\begin{equation}\label{2.75}
  \Fcal_\mathrm{L}[\rho_0] 
  + \int\mathrm{d}\rv\rho_0(\rv)v(\rv).
\end{equation}

\section{Two-stage minimization}
\label{SECtwoStageMinimization}
The essence of the derivation presented in Sec.\ \ref{SEClevy} is a
double minimization of the grand potential functional
(\ref{EQomegaMermin}) of the many-body distribution. In the following,
we spell this out more explicitly. From Sec.\ \ref{SEComegaManyBody}
we know that
\begin{equation}
  \Omega_0 = \min_f \Trcl f \left(H_N-\mu N + \beta^{-1}\ln f\right).
\end{equation}
We decompose the RHS into a double minimization
\begin{equation}
  \Omega_0 = \min_\rho \min_{f\rightarrow\rho} 
  \Trcl f \left(H_N-\mu N + \beta^{-1}\ln f \right),
\end{equation}
where the inner minimization is a search under the contraint the $f$
generates $\rho$ (via relationship (\ref{EQfTorho})).  For
Hamiltonians of the form (\ref{EQhamiltonian}) the above can be
written as
\begin{eqnarray}
  \Omega_0 &=& \min_\rho \min_{f\rightarrow\rho} 
  \Trcl f \left(\sum_{i=1}^N\frac{p_i^2}{2m} + U \right.\nonumber\\ &&\quad\left.
  + \sum_{i=1}^N v(\rv_i)
  -\mu N + \beta^{-1}\ln f \right).
  \label{EQomega0explicitHamiltonian}
\end{eqnarray}
In the expression above
\begin{equation}
  \Trcl f \left(\sum_{i=1}^N v(\rv_i) - \mu N\right) =
  \int d\rv\left(v(\rv)-\mu\right)\rho(\rv),
\end{equation}
because $f\rightarrow\rho$. So we may re-write
(\ref{EQomega0explicitHamiltonian}) as 
\begin{eqnarray}
  \Omega_0 &=& \min_\rho \left\{ \int d\rv \left(v(\rv)-\mu\right)\rho(\rv)
  \right.\\&&\left.\quad
  +\min_{f\rightarrow\rho} 
    \Trcl f\left(\sum_{i=1}^N\frac{p_i^2}{2m}+U+\beta^{-1}\ln f\right)
  \right\}\nonumber
\end{eqnarray}
or 
\begin{equation}
 \Omega_0 = \min_\rho \left\{
  \int d\rv \left(v(\rv)-\mu\right)\rho(\rv) + \Fcal_\mathrm{L}[\rho]
 \right\},
 \label{EQomega0asMinimization}
\end{equation}
where $\Fcal_\mathrm{L}[\rho]$ is given by (\ref{EQFviaLevy}). Clearly
(\ref{EQomega0asMinimization}) is equivalent to (\ref{EQomega0Levy})
and (\ref{EQomegaLevyVariational}).

\section{DFT in the canonical ensemble}
\label{SECcanonical}
One benefit of Levy's method is that it allows straightforward
generalization to the canonical ensemble, as we demonstrate in the
following. In the canonical ensemble (i.e.\ for fixed number of
particles, $N$) the equilibrium many-body distribution functions is
\begin{equation}
  f_{N,0} = Z_0^{-1} \exp(-\beta H_N),
  \label{EQfcanonical}
\end{equation}
where the canonical partition sum is
\begin{equation}
  Z_0 = \Tr_N \exp(-\beta H_N),
\end{equation}
with the canonical trace
\begin{equation}
  \Tr_N = \frac{1}{h^{3N}N!} \int d\rv_1\ldots d\rv_N 
  \int d{\bf p}_1\ldots d{\bf p}_N.
\end{equation}
In analogy to (\ref{EQomegaMermin}) we define the functional
\begin{equation}
  F[f_N] = \Tr_N f_N\left(H_N+\beta^{-1}\ln f_N\right),
  \label{EQfvfunctional}
\end{equation}
where $f_N$ is an arbitrary $N$-body distribution that satisfies
$\Tr_N f_N=1$.  It is easy to show that the (total) Helmholtz free
energy $F_0=-\beta^{-1}\ln Z_0$ is obtained by inserting the
equilibrium distribution (\ref{EQfcanonical}) into the functional
(\ref{EQfvfunctional}), hence
\begin{equation}
  F_0 = F[f_{N,0}].
\end{equation}
Reasoning based on the Gibbs-Bogoliubov inequality, completely
analogous to the arguments presented in Sec.~\ref{SEComegaManyBody},
yields
\begin{equation}
  F_0 = \min_{f_N} F[f_N].
\end{equation}
We decompose this into a double minimization
\begin{equation}
  F_0 = \min_{\rho_N} \min_{f_N\rightarrow\rho_N} F[f_N],
  \label{EQF0doubleMinimization}
\end{equation}
where the canonical one-body density distribution that is generated by
$f_N$ is
\begin{equation}
  \rho_N(\rv)=\Tr_N f_N \hat\rho(\rv),
  \label{EQcanonicalDensity}
\end{equation}
with the density operator $\hat\rho(\rv)$ defined by
(\ref{EQdensityOperator}). Clearly the density defined in this way
satisfies $\int d\rv \rho_N(\rv)=N$, and there are no fluctuations in
the total number of particles. For Hamiltonians of the form
(\ref{EQhamiltonian}), Eq.\ (\ref{EQF0doubleMinimization}) becomes
\begin{eqnarray}
  F_0 &=& \min_{\rho_N} \min_{f_N\to\rho_N} \Tr_N f_N
    \left( \sum_{i=1}^N \frac{p_i^2}{2m} \right.\nonumber\\&&\quad\left.
     + \; U + \sum_{i=1}^N v(\rv_i)
     +\beta^{-1}\ln f_N \right).
\end{eqnarray}
In the above expression
\begin{equation}
  \Tr_N f_N \sum_{i=1}^N v(\rv_i) = \int d\rv v(\rv) \rho_N(\rv),
\end{equation}
because $f_N\to\rho_N$ via (\ref{EQcanonicalDensity}). Hence
\begin{eqnarray}
  F_0 &=& \min_{\rho_N}\left\{
    \int d\rv v(\rv) \rho_N(\rv) \right.\\&&\quad\left.
    +\min_{f_N\to\rho_N} \Tr_N f_N \left(
    \sum_{i=1}^N\frac{p_i^2}{2m} + U + \beta^{-1}\ln f_N \right)
  \right\},\nonumber
\end{eqnarray}
which we write as
\begin{equation}
  F_0 = \min_{\rho_N}\left\{\int d\rv v(\rv)\rho_N(\rv) + F_N[\rho_N]
    \right\},
   \label{EQcanonicalVariationalPrinciple}
\end{equation}
where the intrinsic Helmholtz free energy functional in the canonical
ensemble is defined as
\begin{equation}
  F_N[\rho_N] = \min_{f_N\to\rho_N} \Tr_N f_N \left(
   \sum_{i=1}^N \frac{p_i^2}{2m} + U + \beta^{-1} \ln f_N \right),
\end{equation}
which is formally equivalent to the definition (\ref{EQFviaLevy}) of
$\Fcal_\mathrm{L}$ in the grand ensemble upon identifying the
different traces and different types of many-body distributions. It is
clear that the density distribution $\rho_{N,0}$ that minimizes the
RHS of (\ref{EQcanonicalVariationalPrinciple}) is the true equilibrium
distribution in the canonical ensemble
\begin{equation}
  \rho_{N,0}(\rv) = \Tr_N f_{N,0} \hat\rho(\rv).
\end{equation}
and that $F_0=F_N[\rho_{N,0}]$.
The variational principle (\ref{EQcanonicalVariationalPrinciple})
implies that
\begin{equation}
  \left.\frac{\delta F_N[\rho_N]}{\delta
    \rho_N(\rv)}\right|_{\rho_{N,0}} + v(\rv) =0,
\end{equation}
where the derivative is taken under the constraint $\int d\rv
\rho_N(\rv)=N$.

\section{Discussion and Conclusion}
\label{SECconclusions}
The formulation of DFT rests on the existence and uniqueness of the
intrinsic free energy as a functional of the one-body density for a
given classical system. We have described two methods for defining
this quantity, via Eq.\ (\ref{EQFviaMerminEvans}) based on the
Mermin-Evans argument~\cite{Mermin,Evans2}, and via (\ref{EQFviaLevy})
based on Levy's constrained search~\cite{Levy}. Following the
derivations presented in Secs.\ \ref{SECmerminEvans} and
\ref{SEClevy}, it is clear that these methods are different in
procedure and underlying principles.
 
In the Mermin-Evans sequence of arguments it is formally proved that
the equilibrium many-body probability distribution, $f_0$ is a
functional of the equilibrium one-body density, $\rho_0$. The
existence of this functional rests on a sequence of functional
dependencies. For given interatomic potential $U$ and given one-body
density $\rho$, there is a unique external potential $v$, that
generates this $\rho$. When input into the form of the many-body
distribution in the grand ensemble (\ref{EQf0}), this uniquely
determines $f_0$ as used on the RHS of the definition
(\ref{EQFviaMerminEvans}) of the intrinsic free energy functional
$\Fcal[\rho]$. This chain of dependency is implicit in order to
properly define the free energy functional via
(\ref{EQFviaMerminEvans}). Note that the naive view that the
equilibrium probability distribution, $f_0$, is a function of the
external potential, $v$, such that (\ref{EQFviaMerminEvans}) should
also depend on $v$, gives the impression that functional
(\ref{EQFviaMerminEvans}) is not independent of the external potential
energy. Certainly this is not the case--as one may recall the argument
above (\ref{EQFviaMerminEvans}).

On the other hand, Levy's method does not rely on the above rather
subtle argument. An appealing feature of the constrained search method
is the definition (\ref{EQFviaLevy}) of
$\Fcal_{\mathrm{L}}[\rho]$. Here the intrinsic free energy functional
is explicitly independent of the external potential, which is not as
easily observed from $\Fcal[\rho]$ of
Eq.\ (\ref{EQFviaMerminEvans}). Kohn~\cite{Kohn} and Levy
\cite{levy2010} describe the constrained search method as a two step
minimization procedure, and we have laid out analogous reasoning in
Sec.\ \ref{SECtwoStageMinimization}.

The underlying principle of the Mermin-Evans method of defining the
intrinsic free energy functional is $v$-representability of the trial
density, whereas Levy's functional is based on the weaker condition of
$f$-representability. However, one may restrict the constrained search
to the class of one-body densities that is $v$-representable.  In this
case Levy's functional (\ref{EQFviaLevy}) becomes equal to the
Mermin-Evans functional (\ref{EQFviaMerminEvans}). Hence, the
constrained search method reduces to finding the equilibrium one-body
density which correspond to an (equilibrium) external potential, $v$,
that minimizes functional (\ref{EQomegaLevy}) over all one-body
densities, $\rho$, each associated to a specific $v$. Furthermore,
applying the Legendre transform upon functional (\ref{EQomegaLevy})
and minimizing a set of external potentials which yields a fixed
one-body density, gives Week's free energy functional~\cite{Weeks}.
   
In practice, minimizing Levy's version of the free energy functional
(\ref{EQFviaLevy}) will certainly not be easier than solving the
many-body problem itself. Hence, whether the definition
(\ref{EQFviaLevy}) helps to construct approximations for
grand-canonical free energy functionals remains an open question. For
the case of the canonical ensemble we point the reader to the very
significant body of work that has been carried out to formulate a
computational scheme that permits to capture the effects that arise
due to the constraint of fixed number of particles
\cite{White00prl,White00pre,Hernando02,Gonzalez04jcp,White02,HernandoBlum02}.
While the generalization to equilibrium mixtures is straightforward,
we expect the application of Levy's method to DFT for
quenched-annealed mixtures
\cite{schmidt02pordf,reich04poroned,schmidt02aom,lafuente06qadft} to
constitute an interesting topic for future work.

\begin{acknowledgments}
We thank R.\ Evans, A.J.\ Archer and P.\ Maass for useful
discussions. WSBD acknowledges funding through an Overseas Research
Student Scholarship of the University of Bristol. This work was
supported by the EPSRC under Grant EP/E065619/1 and by the DFG via
SFB840/A3.
\end{acknowledgments}


\begin{thebibliography}{10}

\bibitem{Hohenberg}
P. Hohenberg and W. Kohn, Phys. Rev. {\bf 136},  B 864  (1964).

\bibitem{Mermin}
N.~D. Mermin, Phys. Rev. {\bf 137},  A1441  (1965).

\bibitem{Ebner}
C. Ebner, W.~F. Saam, and D. Stroud, Phys. Rev. A {\bf 14},  2264  (1976).

\bibitem{Ebner2}
W.~F. Saam and C. Ebner, Phys. Rev. A {\bf 15},  2566  (1977).

\bibitem{Ebner3}
C. Ebner and W.~F. Saam, Phys. Rev. Lett. {\bf 38},  1486  (1977).

\bibitem{Evans3}
R. Evans, Adv. Phys. {\bf 28},  143  (1979).

\bibitem{Evans2}
R. Evans,  in {\em Fundamentals of Inhomogeneous Fluids $\mathrm{(ed.\; D.\;
  Henderson)}$} (Dodrecht: Kluwer, New York, 1997), pp.\ 85--175.

\bibitem{Evans4}
R. Evans,  in {\em Les Houches Session XLVIII: Liquids At Interfaces}
  (North-Holland, Amsterdam, 1988), pp.\ 30--66.

\bibitem{Roth}
R. Roth, J. Phys.: Condens. Matter {\bf 22},  063102  (2010).

\bibitem{Hansen}
J.-P. Hansen and I.~R. MacDonald, {\em Theory of Simple Liquids, 3rd Edition}
  (Academic Press (Elsevier), London, 2006).

\bibitem{Levy}
M. Levy, Proc. Natl. Acad. Sci. {\bf 76},  6062  (1979).

\bibitem{Levy2}
P.~W. Ayers and M. Levy, J. Chem. Sci. {\bf 117},  507  (2005).

\bibitem{Kohn}
W. Kohn, Rev. Mod. Phys. {\bf 71},  1253  (1999).

\bibitem{Levy3}
M. Levy, Phys. Rev. A {\bf 26},  1200  (1982).

\bibitem{Coleman}
A.~J. Coleman, Rev. Mod. Phys. {\bf 35},  668  (1963).

\bibitem{Lieb}
E.~H. Lieb, Int. J. Quantum Chem. {\bf 24},  243  (1983).

\bibitem{Levy4}
S.~M. Valone and M. Levy, Phys. Rev. A {\bf 80},  042501  (2009).

\bibitem{levy2010}
M. Levy, Int. J. Quant. Chem. {\bf 110},  3140  (2010).

\bibitem{Weeks}
J.~D. Weeks, J. Stat. Phys. {\bf 110},  1209  (2003).

\bibitem{Chayes}
J.~T. Chayes, L. Chayes, and E.~H. Lieb, Commun. Math. Phys. {\bf 93},  57
  (1884).

\bibitem{White00prl}
J.~A. White, A. Gonz{\'a}lez, F.~L. Rom{\'a}n, and S. Velasco, Phys. Rev. Lett.
  {\bf 84},  1220  (2000).

\bibitem{White00pre}
J.~A. White and S. Velasco, Phys. Rev. E {\bf 62},  4427  (2000).

\bibitem{Hernando02}
J.~A. Hernando, J. Phys.:\ Condensed Matter {\bf 14},  303  (2002).

\bibitem{Gonzalez04jcp}
A. Gonz{\'a}lez, J.~A. White, F.~L. Rom{\'a}n, and S. Velasco, J. Chem. Phys.
  {\bf 120},  10634  (2004).

\bibitem{White02}
J.~A. White and A. Gonz{\'a}lez, J. Phys.:\ Condensed Matter {\bf 14},  11907
  (2002).

\bibitem{HernandoBlum02}
J.~A. Hernando and L. Blum, J. Phys.:\ Condensed Matter {\bf 13},  L577
  (2001).

\bibitem{Widom2}
J.~S. Rowlinson and B. Widom, {\em International Series of Monographs on
  Chemistry: Molecular Theory of Capillarity} (Oxford University Press, Oxford,
  1982).

\bibitem{schmidt02pordf}
M. Schmidt, Phys. Rev. E {\bf 66},  041108  (2002).

\bibitem{reich04poroned}
H. Reich and M. Schmidt, J. Stat. Phys. {\bf 116},  1683  (2004).

\bibitem{schmidt02aom}
M. Schmidt, E. Sch{\"o}ll-Paschinger, J. K{\"o}finger, and G. Kahl,
J. Phys.:\ Condensed Matter {\bf 14},  12099  (2002).

\bibitem{lafuente06qadft}
L. Lafuente and J.~A. Cuesta, Phys. Rev. E {\bf 74},  041502  (2006).

\end{thebibliography}
\end{document}